\documentclass[lettersize,journal]{IEEEtran}

\usepackage[T1]{fontenc}

\usepackage{cite} 
\ifCLASSINFOpdf
  \usepackage[pdftex]{graphicx}
\else
\fi

\usepackage{amsmath}
\usepackage[cmintegrals]{newtxmath}

\usepackage{listings}
\begin{document}

\title{Federated Kalman Filter for Secure IoT-based Device Monitoring Services}

\author{Marc Jayson Baucas,~\IEEEmembership{Student Member,~IEEE,} Petros Spachos,~\IEEEmembership{Senior Member,~IEEE}
\thanks{Marc Jayson Baucas and Petros Spachos are with the School of Engineering, University of Guelph, Guelph, ON N1G2W1, Canada (e-mail: baucas@uoguelph.ca; petros@uoguelph.ca).}}

\maketitle

\begin{abstract}
Device monitoring services have increased in popularity with the evolution of recent technology and the continuously increased number of Internet of Things (IoT) devices. Among the popular services are the ones that use device location information. However, these services run into privacy issues due to the nature of data collection and transmission. In this work, we introduce a platform incorporating Federated Kalman Filter (FKF) with a federated learning approach and private blockchain technology for privacy preservation. We analyze the accuracy of the proposed design against a standard Kalman Filter (KF) implementation of localization based on the Received Signal Strength Indicator (RSSI). The experimental results reveal significant potential for improved data estimation for RSSI-based localization in device monitoring.
\end{abstract}


\section{Introduction}  
The integration of Internet of Things (IoT) devices in monitoring services increased due to their ability to automate and remotely control technologies, usually through a shared wireless network~\cite{smarthome}. IoT devices collect and transmit data to a server and use the information for various remote services, including monitoring and tracking. An example is device monitoring through localization using the Received Signal Strength Indicator (RSSI). The server uses the RSSI data collected from IoT devices to estimate their relative location to the network~\cite{smarthome-loc}. This service allows networks to monitor which devices exist within and around them and ensure that only recognized devices have access.   

However, these monitoring services pose significant challenges to users' data protection~\cite{smarthome-priv}. These services require user data to be effective in accounting for all devices within the network while, at the same time, the devices are prone to potential leakages and theft. To cope with this privacy challenge, in this work, we introduce a platform using Federated Kalman Filter (FKF) with a Federated Learning (FL) approach and blockchain technology to keep the data from users private and protected. These technologies show great potential for the privacy preservation of data\cite{fl-privacy, bc-privacy}, without a significant increase in the system complexity.

\section{Background and Motivation}
\subsection{IoT-Based Device Monitoring}
A device monitoring service is an automated system that monitors the users connected to a network using heterogeneous, wirelessly interconnected devices and sensors~\cite{baucas, smarthome-priv}. It uses the data it collects to automate and regulate the different aspects of the network's environment. The network can then use the information for services such as access control and authorization of users within it~\cite{sh-auth}. 

However, due to issues in wireless communication, device monitoring services pose data security challenges~\cite{sh-security}. Since a large amount of data exchange is required for these services to be effective, the continuous transmission of information across the network creates security vulnerabilities and breaches in user privacy. For instance, device monitoring services use RSSI-based localization to ensure that only trusted devices can access the network within its defined proximity~\cite{smarthome-loc}. However, the RSSI data is directly from the user's IoT devices. This traceable connection introduces a vulnerability and exposes the IoT device to malicious attacks such as data theft and spoofing. As a result, privacy preservation is significant in keeping the system effective while ensuring data is secure within the network~\cite{sh-priv-pres}. 


\subsection{Federated Kalman Filter with Federated Learning}
We selected an FKF with an FL approach to incorporate within the device localization system to ensure the preservation of patient privacy. An FKF is a distributive data fusion and filtering method using Kalman Filtering (KF) as the base~\cite{fed-kf}. A KF is an estimating algorithm for linear systems. KFs are an ideal estimating algorithm for localization data due to the linearity of the distance and localization via RSSI~\cite{rssi-fkf}. FKF maximizes the dynamic estimation of KFs by creating parallel filters that aggregate local results to generate a global model. 

An FKF separates the filtering process globally and locally. The local filter consists of a modified KF that uses values provided by the global filter. It has two steps; the prediction and update steps. The prediction step estimates a system's next state estimate $\widehat{x}_{i(k+1)}$ and covariance values $P_{i(k+1)}$ in its current time index,~$k$. In an FKF implementation, each $\widehat{x}_{i}$ and $P_{i}$ is indexed based on the number of local filters that exist in the system where $i = 1,..., N$, as:
\begin{equation}\begin{split}
  \widehat{x}_{i(k+1)} &= A_{ik}\widehat{x}_{ik} + B_{ik}u_{ik}\\
P_{i(k+1)} &= A_{ik}P_{ik}A_{ik}^{T} + Q_{ik}
\end{split}\label{x-p}\end{equation}

It uses index $i$ and the total number $N$ of the local filter within the system. Also, a system model $A$, measurement model $B$, control input $u$, and noise covariance $Q$. The updating step first solves the Kalman gain $K$ using measurement sensitivity $C$ and measurement error covariance $R$, as:
\begin{equation}
  K_{i(k+1)} = P_{i(k+1)}C_{i(k+1)}^{T}(C_{i(k+1)}P_{i(k+1)}C_{i(k+1)}^{T} + R_{i(k+1)})^{-1}\\
  \label{kgain}
\end{equation}

The resulting values are the local filter's state estimates and covariance values. This finalization process that incorporates the measured input $z$ is:
\begin{equation}
  \begin{split}
    \widehat{x}_{i(k+1)} &= \widehat{x}_{i(k+1)} + K_{i(k+1)}(z_{i(k+1)} - H_{i(k+1)}\widehat{x}_{i(k+1)})\\
    P_{i(k+1)} &= (1-K_{i(k+1)}H_{i(k+1)})P_{i(k+1)}(1-\\
    & K_{i(k+1)}H_{i(k+1)})^{T} + K_{i(k+1)}R_{i(k+1)}K_{i(k+1)}^{T}
  \end{split}
  \label{xpf}
\end{equation} 
   
Usually, this recursive filtering process will result in an estimated representation of the filtered data. However, what makes this KF federated is the additional distributive steps around the prediction stages of the local filter. Instead of using the calculated state estimates $\widehat{x}_{ik}$ and covariance $P_{ik}$ values from the update stage for the following prediction stage calculations, the local KFs will use the values provided by the global filter. This information is defined and divided among the local filters given $\sum_{i=1}^{N, M}\beta_{i} = 1$ where $Q_{ik} = (1/\beta_{i})Q_{k}$, $P_{ik} = (1/\beta_{i})P_{fk}$ and $\widehat{x}_{ik} = \widehat{x}_{fk}$. 

The local filter equations will use these equations to calculate the local state estimates $\widehat{x}_{ik}$ and covariance $P_{ik}$ values. These numbers are then sent to the global filter to obtain the final state estimates $\widehat{x}_{fk}$ and covariance values $P_{fk}$ for the next iteration. First, the global filter calculates its state estimate  $\widehat{x}_{Mk}$ and covariance value $P_{Mk}$. Next, it calculates the final state estimate using the same values as:
\begin{equation}\begin{split}
  P^{-1}_{fk} &= \sum_{i=1}^{N}P^{-1}_{ik} + P^{-1}_{Mk}\\
  \widehat{x}_{fk} &= P_{fk}[P^{-1}_{Mk}\widehat{x}_{Mk} +  \sum_{i=1}^{N}P^{-1}_{ik}\widehat{x}_{ik}]
\end{split}\label{xf-pf}\end{equation}

Our proposed design differs from standard FKFs by introducing an FL approach. The federated aspect of FKFs usually points towards their distributive properties. The FL approach is a machine learning technique using a decentralized strategy to utilize global knowledge for training and tuning its models and filters~\cite{federatedlearning}. Its strength is in effectively preserving the privacy of data. We incorporate it by creating an adaptive loop that enables a real-time KF process within the local filters. Also, we take out the reference signal going to the local filter because, in our FL approach, we adjust global variables to tune the filtering process, similar to training a model based on a known relative distance between the local and the global filters.

\subsection{Private Blockchain}
We integrated blockchain technology to complement the FKF in preserving data privacy within the device localization system. Blockchains are data blocks that are cryptographically linked~\cite{crypto-linked}. We use it as a distributive and tamper-proof ledger that stores and manages historical records of data transactions~\cite{bc-ledger}. Due to its immutability, the blockchain makes it harder to modify the information it holds. Also, its distributive architecture provides multiple backups, reinforcing the confidence in the data structure securing information. Conventionally, there are two blockchains; public and private~\cite{priv-and-pub}. A public blockchain implements a trustless protocol and uses an algorithm to require proof of work (PoW) from devices to compute before it grants them access~\cite{public-bc}. However, this requirement demands high processing power, which is not ideal for IoT devices since they are usually cost-efficient with limited computational and limited capabilities. A private blockchain implements a trusted ledger for consulting pre-defined members when granting access~\cite{private-bc}. However, its size impacts the overall processing speed of the blockchain when managing its users. 

We use a private blockchain in our implementation because it fits our proposed small-scale approach. Since IoT devices are usually low-cost and prioritize power consumption, reducing the processor demand can contribute to the overall sustainability of the device monitoring service within the IoT system. The PoW required by public blockchains can cause complications in implementing the platform due to the latency it might add. Using a private blockchain removes the latency from PoW processing, allowing more leeway for processes from the FKF. Also, since the list is exclusive, the ledger will not be too large to impede the blockchain from carrying out its protocols. There will not be any additional processing required whenever a user attempts to access the server since known users are already within the ledger. With private blockchains, the overall network remains manageable for smaller-scale localization systems. Our proposed platform ensures that the RSSI values do not leave the local filters because only the predictive values go to the global. 

\section{Proposed Platform}\label{meth}
\subsection{Overview}

The proposed platform is an RSSI-based localization implementation using FKF and blockchain technology for IoT-based device monitoring. We aim to address the security issues within these systems due to the potential data leakages in the network. The FKF, with an FL approach, adds a layer of privacy to the network by keeping user data local to the source. This arrangement ensures that information is kept private from the server. As a result, our system preserves data privacy while keeping the localization service effective. The private blockchain will reinforce the distributive network with its cryptographic and tamper-proof features. This addition adds a detection layer by introducing immutability and decentralization, resulting in better defences against malicious attacks.

\begin{figure}[t!]
  \centering
  \includegraphics[width=0.9\linewidth]{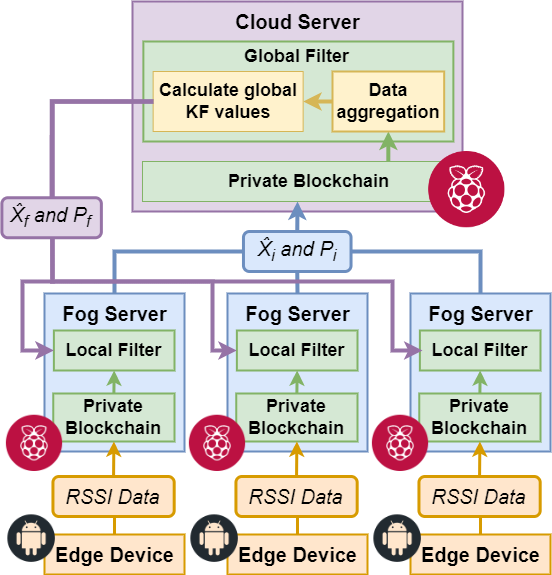}
  \caption{Data flow of proposed RSSI-based localization platform using FKF and private blockchain technology.}
  \label{fkf-diag} 
\end{figure}

The plan is to implement the proposed platform as a low-cost approach for improving the security of RSSI-based localization for IoT-based device monitoring. First, the cloud server contains the global FKF and a copy of the private blockchain. The global filter will manage the filtering parameters the local filters will send. Also, the private blockchain will contain a list of the ID of trusted fog devices, ensuring that only permitted fog devices can transmit data to the cloud. Next, the fog server has the global FKF and a copy of the private blockchain. This device manages the RSSI data collected from the IoT devices. Lastly, the edge devices are the IoT devices that provide the RSSI data to the fog. The local filter uses this data for triangulating these IoT devices and their localization relative to its fog server. A diagram that shows data flowing from the edge to the cloud and a conceptual setup of our implementation is in Fig.~\ref{fkf-diag}.

\subsection{System Components}
There are three design components: the cloud, the fog, and the edge. The cloud includes a laptop with an Intel\textsuperscript{\textregistered}~Core\textsuperscript{\texttrademark}~i7 processor running on Windows 10. This device choice minimizes the impacts of energy consumption on the system by providing a more than capable but portable cloud server to manage the FKF. Next, fog devices use Raspberry Pi 3 B as their central server for device management and RSSI filtering. Since Pis are low-cost, portable, and modular, selecting it as the fog device further reduces the platform's overall energy consumption. Each Pi runs on a Raspian Jesse OS image. Also, we programmed all the scripts it uses using Python 3.6. The edge device is a Google Pixel 6 phone as the source of RSSI data for triangulation and localization. We programmed the FKF and the private blockchain as Python classes that each cloud and fog server initializes. 

The private blockchain class programmed in Python is loaded and initialized within the fog and cloud devices to regulate the data flow and user authorization within the servers. The design of the blockchain implementation contains separate block classes. Each block has part of the list of trusted IoT devices and their IDs. The cloud and fog devices will consult this ledger like a look-up table whenever a device attempts to access or send data. It will manage the data flow if the blockchain acknowledges the device. 

The FKF has two components; the global and local filters. First, we programmed the local filter class within the fog servers. It receives the RSSI data from the edge devices and uses it to triangulate their location through localization. Next, the local filters send the corresponding state estimate and covariance variables to the global filter within the server. The global filter aggregates these values and generates their weights. Finally, it sends these values to each local filter for the following filtering iteration. 

\section{Experimental Results}\label{result}
\subsection{Testbed}
We designed a testbed to examine our proposed RSSI-based localization platform. It compares our FKF design against standard KFs to determine its ability to keep data filtering consistent and accurate while preserving data privacy. The metrics we used to measure the accuracy are localization reliability and prediction precision using Root Mean Squared Error (RMSE) and the RSSI prediction accuracy. We calculated the RMSE through the following equation:
\begin{equation}
  RMSE = \sqrt{\frac{\sum_{i=1}^{n} (Predicted_{i} - Obeserve_{i})^{2}}{n}}
  \label{rmse}
\end{equation}
Also, we calculated the RSSI accuracy through a percent accuracy formula defined as:  

\begin{equation}
  RSSI_{acc} = \left(1 - \left|\frac{Theoretical - Measured}{Theoretical}\right|\right) * 100\%
  \label{per_acc}
\end{equation}

The lower the RMSE and the higher the RSSI prediction accuracy, the more accurate the filter's estimation. We chose localization reliability and prediction precision because these can evaluate and measure our proposed design's ability to filter RSSI data precisely and consistently localize devices. Also, we further analyze the performance of the proposed method by comparing the computational complexity. We arranged our testbed to have four local filters and a global filter. The global filter is within the central server of the platform. The local filters within fog devices will receive the RSSI values via WiFi signal strength from the edge devices. We situated these fog servers around a room as triangulation anchors for the localization system. The edge device is placed close to this perimeter at a known distance to provide the RSSI for analyzing the KF precision.

\subsection{Performance of Global Filter compared to Local Filter}
\begin{figure}[t!]
  \centering
  \includegraphics[width=\linewidth]{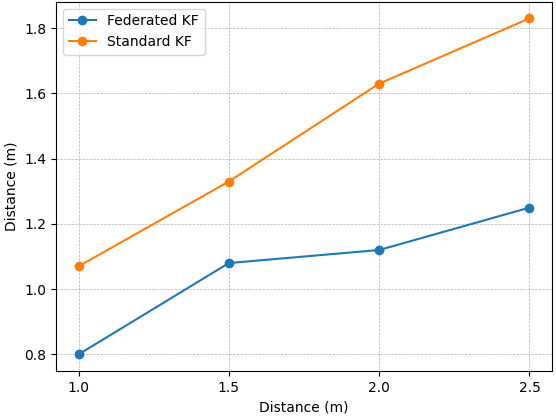}
  \caption{RMSE calculation of FKF and SKF configurations at known distances between the fog servers and the edge device.}
  \label{fkf-rmse} 
\end{figure}

This experiment has two configurations: the FKF and the standard KF. The FKF configuration is our proposed platform, while the standard KF configuration does not have the central server to send its parameters. All KFs use a path loss factor of 2.00 and a system loss constant of 57 for the distance calculations. Also, the known distances between the edge and fog devices are 1, 1.5, 2, and 2.5 meters. We take the average RMSE to represent the localization reliability of the configuration for each known distance. The plot showing the RMSE values from each experiment iteration is in Fig.~\ref{fkf-rmse}. We can observe through it that the RMSE values of the FKF method were lower overall compared to the standard KF. This observation suggests that the FKF has a more consistent and reliable localization filter. Also, it shows that the proposed method is more capable of accounting for spikes in RSSI. 

Meanwhile, the average accuracy of the standard KF and FKF was 89.85\% and 87.56\%, respectively. We can attribute the higher RSSI from the standard KF to the global KF having to aggregate data from the local KFs causing the weights to affect the prediction process. This calculated accuracy is relative to the measured value. So, a slightly higher percentage does not mean a significantly accurate filter. Considering the volatility of RSSIs, the FKF maintaining a close predictive accuracy with the standard KF, even with the added weight aggregation, further reinforces its feasibility. Also, the lower RSME suggests a less consistent filter when dealing with sudden spikes in the RSSI data. Although the estimates of the KF are slightly closer to the measurements, its localization is less stable. Also, a 2\% difference is insignificant for RSSIs since they are always whole numbers within the 50-60~dBm range. Therefore, these results present the FKF as a more reliable and equally precise option.   

In terms of complexity, we can present the computational costs of a standard KF as $O(n_{m}^{3} + n_{p}^{2})$ considering the measurement $n_{m}$ and prediction $n_{p}$ phases. The complexity of the measurement phase is higher due to more matrix inversions. The localization process $n_{l}$ goes through each predicted value. So, it increases the cost to $O(n_{m}^{3} + n_{p}^{2} + n_{l}^{2})$. For the FKF, it is the same. However, we move the prediction stages to the global filter. The result is a complexity that we can split into $O_{local}(n_{m}^{3} + n_{l}^{2}) + O_{global}(n_{p}^{2})$. Also, the assigning and calculating weights $n_{w}$ is carried each for each prediction, increasing the complexity to $O_{local}(n_{m}^{3} + n_{l}^{2}) + O_{global}(n_{p}^{2} + n_{w}^{2})$. We can observe that even with the reallocation and added processes, the overall computational costs do not change. Also, the added global filter has a more capable processor, which lowers the impact of its computational costs on the overall method.  

\section{Conclusion}\label{conc}
The proposed platform combines an FKF with an FL approach and a private blockchain with RSSI-based localization for device monitoring services. It introduces a security layer that ensures privacy preservation through the FKF and FL combination and access authorization through the private blockchain within the monitoring service. We evaluated our proposed design's localization reliability and prediction precision against a standard KF using RMSE and RSSI prediction accuracy. Also, we discussed the computational costs of each method. Each evaluation investigated if the service's integrity is maintained even after adding the FL process. With the FL and blockchain adding security, we observed better overall accuracy from the proposed RSSI-based localization system.   

\bibliographystyle{IEEEtran}
\bibliography{IEEEabrv,fedlocbib}
\end{document}